\documentclass[conference]{IEEEtran}

\usepackage{cite}
\usepackage[pdftex]{graphicx}
\graphicspath{{fig/}}
\usepackage{amsmath}
\usepackage{array}
\usepackage[caption=false,font=footnotesize]{subfig}
\usepackage{fixltx2e}
\usepackage{url}

\usepackage{lipsum}
\usepackage{bbm} 
\usepackage{amssymb} 
\usepackage{color,soul} 
\usepackage{booktabs} 
\usepackage[ruled]{algorithm2e} 
\usepackage{multirow} 
\usepackage{tikz} 
\usetikzlibrary{positioning}
\usetikzlibrary{shapes,arrows,shadows}
\usetikzlibrary{decorations.pathreplacing}
\usepackage[normalem]{ulem} 
\usepackage{adjustbox}

\makeatletter
\def\ps@IEEEtitlepagestyle{%
  \def\@oddfoot{\mycopyrightnotice}%
  \def\@evenfoot{}%
}
\def\mycopyrightnotice{%
  {\footnotesize\hfill Accepted for IEEE GLOBECOM 2017. \copyright~2017 IEEE\hfill}%
  \gdef\mycopyrightnotice{}
}
\makeatother

\begin{document}
\title{Modeling a Traffic Remapping Attack Game in a Multi-hop Ad Hoc Network}
\author{\IEEEauthorblockN{Jerzy Konorski}
\IEEEauthorblockA{Faculty of Electronics, \\Telecommunications and Informatics\\
Gdansk University of Technology\\
Gdansk, Poland\\
Email: jekon@eti.pg.gda.pl}
\and
\IEEEauthorblockN{Szymon Szott}
\IEEEauthorblockA{AGH University of Science and Technology\\
Faculty of Computer Science, \\Electronics and Telecommunications\\
Krakow, Poland\\
Email: szott@kt.agh.edu.pl}}

\maketitle

\begin{abstract}
In multi-hop ad hoc networks, selfish nodes may unduly acquire high quality of service (QoS) by assigning higher priority to source packets and lower priority to transit packets. Such traffic remapping attacks (TRAs) are cheap to launch, impossible to prevent, hard to detect, and harmful to non-selfish nodes. While studied mostly in single-hop wireless network settings, TRAs have resisted analysis in multi-hop settings. In this paper we offer a game-theoretic approach: we derive a formal model of opportunistic TRAs, define a TRA game with a heuristic rank-based payoff function, and propose a boundedly rational multistage attack strategy that both selfish and non-selfish nodes are free to use. Thus non-selfish nodes are allowed to respond in kind to selfish ones. We characterize the form of equilibrium that the multistage play reaches and verify via simulation that it often coincides with a Nash equilibrium in which harmful TRAs are curbed in the first place, whereas harmless ones need not be.

\end{abstract}

\section{Introduction}

Nodes forming an ad hoc network may abuse the network's mechanisms to achieve an undue increase of the received quality of service (QoS). 
E.g., by disseminating false routing advertisements, a node may prevent establishing paths that traverse it and thus avoid forwarding transit traffic. This necessitates secure routing protocols \cite{Kannhavong2007}, intrusion detection systems \cite{Hassanzadeh2016}, or trust management frameworks \cite{Movahedi2016}.
However, a more subtle method exists, referred to as the \emph{traffic remapping attack} (TRA), that can be used to bring an attacker better QoS at a low execution cost and a low risk of detection.
A node performing a TRA falsely assigns traffic to classes: either source packets are assigned higher priority or transit packets are assigned lower priority (or both).

While TRAs are applicable to any network with class-based traffic differentiation, in ad hoc networks using IEEE 802.11 they rely on the enhanced distributed channel access (EDCA) function. 
EDCA defines four access categories (ACs), each with its own set of medium access parameters to determine the probability and duration of channel access. 
Packets are mapped to ACs based on the Distributed Services Code Point (DSCP) set in their IP header, which in turn is based on the traffic's Class of Service (CoS) \cite{Stankiewicz2011}. 
The CoS-to-DSCP mapping is done according to administrator policies, while the DSCP-to-AC mapping is implemented by network-layer packet mangling software.
Thus, software such as Linux \texttt{iptables} are enough to execute a TRA (which amounts to a false DSCP-to-AC mapping). 
This is in contrast to other selfish attacks, such as medium access parameter modification, which require tampering with the wireless card drivers.
Furthermore, TRAs are difficult to detect: determining if the monitored higher-layer traffic matches its class designation requires deep packet inspection \cite{Konorski2014}.

TRAs have so far mostly been studied in single-hop (ad hoc or infrastructure-based) wireless network settings \cite{Li2005,Ghazvini2013,Konorski2014,Politis2016,Konorski2017}, where they have been observed to drastically reduce the throughput of non-selfish nodes \cite{Konorski2014}.
In multi-hop settings such as mobile ad hoc networks (MANETs), the security threat posed by TRAs is aggravated by their multi-hop impact: once assigned false priority, a packet retains its QoS designation further down the path to the destination. Additionally, a selfish node can both promote its source traffic and demote transit traffic \cite{Szott2014}.
An introductory study of the impact of TRAs and suitable countermeasures for multi-hop wireless settings can be found in \cite{Szott2017}.
In this paper, we perform a systematic analysis using a game-theoretic approach and providing the following contributions:
\begin{itemize}
\item from a model of quasi-static traffic patterns in an IEEE 802.11 EDCA-based mobile ad hoc network (MANET) (Section~\ref{sec:model}), we derive a formal model of \emph{opportunistic} TRAs (Section~\ref{sec:attacks}),
\item next, in Section~\ref{sec:performance} we devise a heuristic rank-type end-to-end performance metric to quantify the cost of TRAs both for attacker and non-attacker nodes,
\item using the cost metric as a payoff function, in Section~\ref{sec:game} we formally define and characterize a single-stage \emph{TRA game} in which both selfish and nonselfish nodes are free to launch TRAs, hence the latter may defend themselves against TRAs by responding in kind,
\item we argue that MANET nodes are likely to exhibit \emph{bounded rationality} \cite{Ho1996}, i.e., limited complexity, perseverance, and foresight; for such nodes in Section~\ref{sec:multistage-game} we propose a multistage TRA strategy and verify experimentally that it reaches a form of equilibrium,
\item we verify that the multistage play most often ends up at, or close to, a Nash equilibrium of the single-stage TRA game, whereupon both attacker and non-attacker nodes typically benefit from TRAs.
\end{itemize}
Section~\ref{sec:conclusions} concludes the paper and outlines some directions of future work.

\section{MANET Model}
\label{sec:model}
Let $G=\langle N,L \rangle$ be a directed graph representing the current (quasi-static) MANET hearability topology, where $N$ is the set of nodes, $L \subset N \times N$, and $(i, j) \in L$ iff $i \neq j$ and $j$ is in the hearability range of $i$. By $N^*$ we denote the set of all directed acyclic routes in $G$ (i.e., sequences of nodes such that for each two consecutive nodes $i$ and $j$, $(i, j) \in L$). 

Let $R \subseteq N^*$ be the set of end-to-end routes in $G$ as determined by the routing algorithm in use. 
Each $r \in R$ is represented as a sequence $(s_r, \dots, d_r)$ of involved nodes, where $s_r$ and $d_r$ are source and destination nodes.
We adopt the following notation for routes: write $i \in r$ if $r$ involves node $i$; for $i, j \in r$ write $i <_r j$ ($i \leq_r j)$ if $i$ precedes (precedes or coincides with) $j$ on $r$; for $i \in r \setminus \{d_r\}$ denote by $succ_{r,i}$ the immediate successor of $i$ on $r$, and for $i \in r \setminus \{s_r\}$ define $pred_{r,i}$ as the immediate predecessor of $i$ on $r$ ($pred_{r,s_r}$ is defined as $s_r$). Let $P_{r,i }= \{j | j \leq_r i\}$ be the set of nodes that precede or coincide with $i$ on $r$.

Further assume that MANET traffic is composed of \emph{end-to-end (e2e-) flows}, each of which is a collection of packets of the same $class \in CoS$ and moving along the same route.
The corresponding MAC-layer frames are assigned ACs, which they carry in the AC fields contained in their headers, and handled accordingly by EDCA.
Let $AC$ be the set of distinguished ACs. For ease of presentation we restrict the used ACs to \emph{VO} (assigned to voice traffic) and \emph{BE} (assigned to best-effort traffic), i.e., $AC = \{VO, BE\}$, with \emph{VO} enjoying (statistical) priority over \emph{BE} at the MAC layer.

Since packet mangling software in fact amounts to a CoS-to-AC mapping, one can define a function $mang: CoS \to AC$ such that $mang(class)$ is the AC that the class of service $class \in CoS$ should map to.
Then, an e2e-flow is represented as $(r, ac)$, where $r \in R$ is its route and $ac = mang(class) \in AC$ is its intrinsic AC as returned at $s_r$.  Let $F \subseteq R \times AC$ be the (quasi-static) set of e2e-flows offered by MANET users. Without loss of generality we assume that at least one e2e-flow is offered at each node, i.e., $\{s_r | (r, ac) \in F\} = N$. 

We refer to \emph{hop (h-)flows} as the granulation level at which traffic is recognized at a next-hop node.
At a given $i \in r$, packets of e2e-flow $(r, ac)$ transmitted by $j=pred_{r,i}$, whose frame headers contain AC fields with $hac \in AC$, are recognized as an h-flow $(j, r, hac)$  (in general, it may be that $hac \neq ac$, since the AC fields can be modified hop-by-hop). For completeness, assume that e2e-flow $(r, ac)$ is recognized at $s_r$ as $(s_r,r,ac)$. Let $H \subseteq N \times R \times AC$ be the set of recognizable h-flows.

Autonomous operation of node $i$ is modeled as $map_i: H \rightarrow AC$. For an incoming h-flow $(j, r, hac)$ recognized at node $i$, where $j = pred_{r,i}$ and $i \in r \setminus \{s_r, d_r\}$, $map_i(j, r, hac)$ is the new AC field transmitted by $i$ further along $r$.

\section{Attack Model}
\label{sec:attacks}
A \emph{traffic remapping attack} (TRA) that a node $i \in r$ launches upon an incoming h-flow $(j, r, hac)$, where $j = pred_{r,i}$ and $i \in r \setminus \{d_r\}$, consists in configuring $map_i(j, r, hac) \neq hac$. Such a definition captures the fact that the setting of AC fields under a TRA is both perfectly legal (in that the use of $map_j$ is correct) and ill-willed (inconsistent with \emph{mang}). In light of this, behavior of node $i$ with respect to h-flow $(j, r, hac)$ can be classified as (i) \emph{neutral}, if $map_i(j, r, hac) = hac$, (ii) \emph{upgrading TRA} (TRA$^+$), if $hac = BE$ and $map_i(j, r, BE) = VO$, or (iii) \emph{downgrading TRA} (TRA$^-$) if $hac = VO$ and  $map_i(j, r, VO) = BE$. Node behavior is moreover assumed \emph{plausible} in that it never downgrades source traffic or upgrades transit traffic, i.e., $map_i(j, r, hac) = hac$ if $(hac = VO$ and $i = s_r)$ or $(hac = BE$ and $i \neq s_r)$. Nodes that exhibit TRA behavior will be called \emph{attackers} and their set will be denoted $A$. 
Let us assume that each attacker is \emph{opportunistic}, i.e., launches a TRA$^+$ or a TRA$^-$ upon all h-flows it recognizes, subject to the plausibility constraints.

Note that with respect to a given e2e-flow $(r, ac)$, a plausible opportunistic attacker $i$ does not actually modify any AC field when $i \not\in r$ or $i = d_r$ (in the latter case $i$ will behave neutrally), or when $ac = BE$ and $i \neq s_r$, or, finally, when $ac = VO$ and $i \neq s_r$ and $i$ recognizes the e2e-flow as $(j, r, BE)$, where $j = pred_{r,i}$ (i.e., when a TRA$^-$ has been launched by one of the nodes preceding $i$ on $r$).

\begin{table}[tbp]
  \footnotesize
  \centering
  \caption{Example of a 10-node MANET with flow-sparse traffic}
  \subfloat[Hearability topology (node incidence matrix)]{%
\begin{tabular}{@{}c|cccccccccc@{}}
   & 1 & 2 & 3 & 4 & 5 & 6 & 7 & 8 & 9 & 10 \\ \midrule
1  & - & 0 & 1 & 1 & 0 & 1 & 0 & 0 & 0 & 1 \\
2  & 0 & - & 1 & 0 & 1 & 0 & 0 & 1 & 0 & 0 \\
3  & 1 & 1 & - & 0 & 1 & 0 & 1 & 1 & 1 & 1 \\
4  & 1 & 0 & 0 & - & 1 & 1 & 1 & 1 & 0 & 0 \\
5  & 0 & 1 & 1 & 1 & - & 0 & 0 & 1 & 1 & 0 \\
6  & 1 & 0 & 0 & 1 & 0 & - & 0 & 1 & 0 & 0 \\
7  & 0 & 0 & 1 & 1 & 0 & 0 & - & 1 & 1 & 1 \\
8  & 0 & 1 & 1 & 1 & 1 & 1 & 1 & - & 0 & 0 \\
9  & 0 & 0 & 1 & 0 & 1 & 0 & 1 & 0 & - & 0 \\
10 & 1 & 0 & 1 & 0 & 0 & 0 & 1 & 0 & 0 & -
\end{tabular}%
  }\vspace{0.2cm}
  \subfloat[TRAs experienced by e2e flows for $A = \{1, 3,8,9\}$]{%
\begin{tabular}{@{}llll@{}}
\toprule
Flow & $r$             & $ac$ & TRA                     \\ \midrule
1  & 1  3  9 5  4 & \emph{VO} & TRA$^-$ at 3               \\
2  & 2 8  3       & \emph{BE} &                         \\
3  & 3 10  7 4  5 & \emph{VO} &                         \\
4  & 4 8          & \emph{BE} &                         \\
5  & 5 8  6  1 10 & \emph{VO} & TRA$^-$ at  8              \\
6  & 6 8          & \emph{BE} &                         \\
7  & 7 8  6  4  5 & \emph{VO} & TRA$^-$ at  8              \\
8  & 8 3  2       & \emph{BE} & TRA$^+$ at 8 \& TRA$^-$ at  3 \\
9  & 9 7  3  5  2 & \emph{VO} & TRA$^-$ at  3              \\
10 & 10 1  4      & \emph{BE} &                         \\
\bottomrule
\end{tabular}%
  }\label{tab:example-sparse}
\end{table}

The example in Table~\ref{tab:example-sparse}a shows a 10-node MANET with $G$ represented as a hearability topology (node incidence matrix) and with a given set $A = \{1,3,8,9\}$ of opportunistic attackers.
Routes of e2e-flows were of uniformly distributed lengths $2\dots 5$ and were selected at random\footnote{The selected random routes were not necessarily the shortest paths.}.
Each node is a source of one e2e-flow, half of the flows being \emph{VO} (we will refer to this traffic pattern as \emph{flow-sparse}).
For each e2e-flow it is indicated what TRAs have been experienced and at which node.

The following can be observed in Table~\ref{tab:example-sparse}b regarding the selected e2e-flows:

\begin{itemize}
\item e2e-flow \#3 with $ac=VO$ has an attacker source, which, however, does not launch a TRA$^{-}$ due to the plausibility constraints,

\item no e2e-flow with $ac=VO$ has an attacker destination, but even if it did, the destination would behave neutrally due to the plausibility constraints,

\item for the same reason, e2e-flows \#2,4, and 6 with $ac=BE$ are not attacked at their attacker destinations, e2e-flow \#10 is not attacked at an attacker forwarder (node 1), which could only launch a TRA$^{+}$, and e2e-flow \#1 is not attacked at its attacker source (node 1),

\item e2e-flows \#1 and 5 with $ac = VO$ each encounter two attacker forwarders, of which the first launches a TRA$^{-}$, hence the second no longer has to,

\item e2e-flow \#8 experiences a combination of a TRA$^{+}$ at its source node 8 and a TRA$^{-}$ at node 3; this is the maximum number of attacks an e2e-flow can experience. 
\end{itemize}
Note that if all the nodes were opportunistic attackers ($A=N$) then all e2e-flows with $|r|>2$ would be recognized at destination as \emph{BE} h-flows.

\begin{table}[tbp]
  \footnotesize
  \centering
  \caption{Example of a 10-node MANET with flow-dense traffic}
  \subfloat[Hearability topology (node incidence matrix)]{%
\begin{tabular}{@{}c|cccccccccc@{}}
   & 1 & 2 & 3 & 4 & 5 & 6 & 7 & 8 & 9 & 10 \\ \midrule
1  & - & 0 & 1 & 0 & 1 & 1 & 0 & 1 & 0 & 1 \\
2  & 0 & - & 1 & 1 & 0 & 1 & 0 & 1 & 0 & 1 \\
3  & 1 & 1 & - & 1 & 1 & 1 & 1 & 1 & 1 & 1 \\
4  & 0 & 1 & 1 & - & 1 & 0 & 1 & 1 & 0 & 1 \\
5  & 1 & 0 & 1 & 1 & - & 1 & 1 & 1 & 1 & 1 \\
6  & 1 & 1 & 1 & 0 & 1 & - & 0 & 1 & 1 & 0 \\
7  & 0 & 0 & 1 & 1 & 1 & 0 & - & 1 & 0 & 1 \\
8  & 1 & 1 & 1 & 1 & 1 & 1 & 1 & - & 0 & 0 \\
9  & 0 & 0 & 1 & 0 & 1 & 1 & 0 & 0 & - & 0 \\
10 & 1 & 1 & 1 & 1 & 1 & 0 & 1 & 0 & 0 & -
\end{tabular}%
  }\vspace{0.2cm}
  \subfloat[TRAs experienced by e2e flows for $A = \{1, 3,8,9\}$]{%
\begin{tabular}{@{}llll@{}}
\toprule
Flow & $r$             & $ac$ & TRA                     \\ \midrule
1  & 1  3  9 5  4  & \emph{VO} & TRA$^-$ at 3               \\
2  & 1 8  3        & \emph{BE} & TRA$^+$ at 1 \& TRA$^-$ at  8 \\
3  & 2 10  7 4  5  & \emph{VO} &                         \\
4  & 2 8           & \emph{BE} &                         \\
5  & 3 8  6  1 10  & \emph{VO} & TRA$^-$ at  8              \\
6  & 3 8           & \emph{BE} & TRA$^+$ at  3              \\
7  & 4 7  8  6  5  & \emph{VO} & TRA$^-$ at  8              \\
8  & 4 3  2        & \emph{BE} &                         \\
9  & 5 7  3  2  4  & \emph{VO} & TRA$^-$ at  3              \\
10 & 5 1           & \emph{BE} &                         \\
11 & 6 3  9        & \emph{VO} & TRA$^-$ at  3              \\
12 & 6 1           & \emph{BE} &                         \\
13 & 7 5  8  1  3  & \emph{VO} & TRA$^-$ at  8              \\
14 & 7 8  5        & \emph{BE} &                         \\
15 & 8 4 10  3  2  & \emph{VO} & TRA$^-$ at  3              \\
16 & 8 2  6        & \emph{BE} & TRA$^+$ at  8              \\
17 & 9 5  8  7 10  & \emph{VO} & TRA$^-$ at  8              \\
18 & 9 6           & \emph{BE} & TRA$^+$ at  9              \\
19 & 10 5  3  6  2 & \emph{VO} & TRA$^-$ at  3              \\
20 & 10 4          & \emph{BE} &                       \\ \bottomrule
\end{tabular}%
  }\label{tab:example-dense}
\end{table}

A similar example, where each node is a source of two e2e-flows, one \emph{VO} and one \emph{BE} (which we will refer to as the \emph{flow-dense} traffic pattern) is presented in Table~\ref{tab:example-dense}. 
Under such a pattern, typically a larger proportion of flows are attacked (either experience a single TRA or a combination of a TRA$^+$ and a TRA$^-$ along $r$).

\section{Performance under TRAs}
\label{sec:performance}
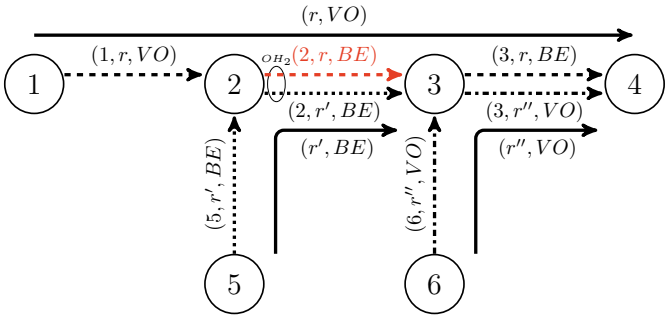
\begin{figure}[tbp]
\vspace{0.061in}
\centering
\tikzstyle{every pin edge}=[-]
\begin{adjustbox}{width=\columnwidth}
\begin{tikzpicture}[->,>=stealth',shorten >=1pt,auto,node distance=3.4cm,
                    thick,main node/.style={minimum width=1cm,circle,draw,font=\sffamily\bfseries\Large}]

  \definecolor{myred}{RGB}{229,63,38}
  \definecolor{myorange}{RGB}{0,0,0}

  \node[main node] (1) {$1$};  
  \node[main node] (2) [right of=1] {$2$};
  \node[main node] (3) [right of=2] {$3$};
  \node[main node] (4) [right of=3] {$4$};
  \node[main node] (5) [below of=2] {$5$};
  \node[main node] (6) [below of=3] {$6$};  	

  \coordinate[right=.2cm of 2] (e1);	
  \draw[thin] (e1) circle [x radius=0.15cm, y radius=.3cm] node [above=.2cm,font=\tiny] {$OH_2$};

  \draw[ultra thick,transform canvas={yshift=2ex},->] (1.north) -- (4.north) node[above,midway] {$(r,VO)$};

  \coordinate[right=.7cm of 5.north] (t1);	
  \coordinate[above=2.1cm of t1] (t2);
  \coordinate[right=2.1cm of t2] (t3);	
  \draw[rounded corners,ultra thick,->] (t1) -- (t2) -- (t3) node[below,midway] {$(r',BE)$};

  \coordinate[right=.7cm of 6.north] (t4);	
  \coordinate[above=2.1cm of t4] (t5);
  \coordinate[right=2.1cm of t5] (t6);	
  \draw[rounded corners,ultra thick,->] (t4) -- (t5) -- (t6) node[below,midway] {$(r'',VO)$};

  \draw[ultra thick,dashed,->,transform canvas={yshift=1ex},myorange] (1)--(2) node[above,midway]  {$(1,r,VO)$};
  \draw[ultra thick,dashed,->,transform canvas={yshift=1ex},myred] (2)--(3) node[above,midway] {$(2,r,BE)$};
  \draw[ultra thick,dotted,->,transform canvas={yshift=-1ex},myorange] (2)--(3) node[below,midway] {$(2,r',BE)$};
  \draw[ultra thick,dashed,->,transform canvas={yshift=1ex},myorange] (3)--(4) node[above,midway]  {$(3,r,BE)$};
  \draw[ultra thick,dashdotted,->,transform canvas={yshift=-1ex},myorange] (3)--(4) node[below,midway] {$(3,r'',VO)$};
  \draw[ultra thick,dotted,->,myorange] (5)--(2) node[above,midway,rotate=90] {$(5,r',BE)$};
  \draw[ultra thick,dashdotted,->,myorange] (6)--(3) node[above,midway,rotate=90] {$(6,r'',VO)$}; 
\end{tikzpicture}
\end{adjustbox}
\caption{Illustration of e2e-flows (solid arrows), h-flows (patterned lines), and related concepts in a grid-shaped MANET with $A=\{2\}$. The attacker performs a TRA$^-$ on e2e-flow $(r,VO)$, hence $hac_2(r,VO)=BE$. The set of two outgoing h-flows at node $2$ is labeled $OH_2$. For the featured h-flow $(2,r,BE)$ (red), all other h-flows belong to the set of competing h-flows $CH_2(r,VO)$. 
Note that the attacked flow receives worse QoS on all hops following node $2$ (i.e., is forwarded as BE).}
\label{fig:model}
\end{figure}

For an e2e-flow $(r,ac) \in F$ denote by $hac_i(r,ac)$ the AC of the flow's packets transmitted by node $i \in r \setminus \{d_r\}$ and received at node $succ_{r,i}$. This AC is returned by the superposition of all $map_j$, $j \leq_r i$, and the corresponding h-flow can be designated as $(i, r, hac_i(r, ac))$.  
Fig.~\ref{fig:model} illustrates e2e-flows, h-flows, and related notions.
Given the set $A$ of opportunistic attackers, one can derive $hac_i(r, ac)$ as follows: 

\begin{align}
\label{hacBE}
&hac_i(r,BE)=\left\{
\begin{aligned}
& VO,\quad s_r\in A\wedge P_{r,i} \setminus \{s_r\} \cap A=\emptyset \\
& BE,\quad \text{otherwise}
\end{aligned} \right. ,\\
&hac_i(r,VO)=\left\{
\begin{aligned}
\label{hacVO}
& VO,\quad P_{r,i} \setminus \{s_r\} \cap A=\emptyset \\
& BE,\quad \text{otherwise}
\end{aligned} \right. .
\end{align}
That is, $hac_i(r, BE) = VO$ if a TRA$^+$ has been launched at the source and no TRA$^-$ has been launched by the time the flow's packets reach $i$, and $hac_i(r, BE) = BE$ if no TRA or both a TRA$^+$ and a TRA$^-$ have been launched.
Similarly, $hac_i(r, VO) = VO$ if no TRA$^-$ has been launched at nodes other than the source, and $hac_i(r, VO) = BE$ if a TRA$^-$ has been launched.

The set of outgoing h-flows at node $i$ is
\begin{equation}
\label{OH}
OH_i = \{(i, r, hac_i(r, ac)) | (r, ac) \in F \wedge i \in r \setminus \{d_r\}\}.
\end{equation}
Important from the viewpoint of an outgoing h-flow $(i, r, hac_i(r, ac))$ is the set of \emph{competing h-flows}, i.e., h-flows it has to locally compete with for local wireless spectrum.
These are: (a) other outgoing h-flows at $i$, which compete via the local transmission queue, (b) outgoing h-flows at nodes in the hearability range of $i$, which compete via CSMA/CA, and (c) outgoing h-flows at nodes in the hearability range of $succ_{r,i}$ but not of $i$ (i.e., hidden from $i$), which compete via exclusive-OR reception at $succ_{r,i}$:
\begin{multline}
CH_i(r,ac) = OH_i \setminus \{(i,r,hac_i(r,ac))\} \cup \\
 \bigcup\nolimits_{j:(j,i)\in L}OH_j
\cup \bigcup\nolimits_{j:(j,succ_{r,i})\in L \land (j,i) \not\in L}OH_j.
\end{multline} 
For the above h-flow, the pair $[hac, CH]_i(r, ac)$ determines per-hop performance at node $i$, where we use a succinct notation $[a, b]_i(x)$ instead of $[a_i(x), b_i(x)]$. We propose a rank-type per-hop performance metric $rank_i(r, ac)$ reflecting that an h-flow is better off at a node if it is \emph{VO} and competes with fewer (and preferably \emph{BE}) h-flows. Accordingly, the metric should rank the vectors $[hac,vo,be]_i(r, ac)$, where $vo_i(r, ac)$ and $be_i(r, ac)$ represent the number of \emph{VO} and \emph{BE} h-flows in $CH_i(r, ac)$.

To validate $rank(\cdot)$, we have used the Markovian model of EDCA \cite{szott2010ieee} to calculate the normalized per-hop saturation throughput $S_i(hac, r, ac)$ of e2e-flow $(r,ac)$ at node $i$, given $hac=hac_i(r,ac) \in \{VO,BE\}$, and $vo=vo_i(r, ac)$ and $be=be_i(r, ac)$ each ranging from 0 to 10. For the resulting $2 \cdot 11 \cdot 11(2 \cdot 11 \cdot 11-1)/2$ pairs of throughput values, $rank(\cdot)$ represents a good fit if
\begin{multline}
\label{implic}
rank_i(hac, r, ac) \le rank_i(hac', r', ac')
\\
\text{iff}~S_i(hac, r, ac)>S_i(hac', r', ac')
\end{multline}
holds for a high percentage of pairs. (Obviously, a small rank is desirable.) A heuristic metric is
\begin{multline}
\label{rank}
rank_i(hac, r, ac)=\mathbbm{1}_{hac=BE} \cdot \alpha \cdot (vo+\mathbbm{1}_{vo>1 \lor be>2})
\\
+ \beta \cdot (vo+\mathbbm{1}_{hac=BE})+be,
\end{multline}
where $\mathbbm{1}_x = 1$ if logical condition $x$ is true and 0 otherwise, and the best fit (99.13\%) occurs at $\alpha =40$ and $\beta =10$. The preferences of h-flows are reflected in that $vo$ has more impact upon $rank_i(hac,r, ac)$ than does $be$ (since $\beta > 1$), and there is distinct separation between $hac=VO$ and $hac=BE$ (since $\alpha \gg \beta$). 



Our rank metric induces a heuristic e2e-flow cost metric we call $flowcost$, additive for \emph{VO} traffic delay and bottleneck-type for \emph{BE} traffic throughput (to make both metrics comparable in magnitude, we take per-node \emph{VO} traffic delay):
\small
\begin{equation}
flowcost_{(r,ac)}(A)=
\left\{
\begin{aligned}
& \frac{\sum_{i\in r\setminus\{d_r\}} rank_i(hac,r,ac)}{|r|-1}, &ac=VO \\
& \max_{i\in r\setminus\{d_r\}}rank_i(hac,r,ac), &ac=BE
\end{aligned} \right. ,
\end{equation}
\normalsize
where $hac$ is given by (\ref{hacBE}) and (\ref{hacVO}). (Since $hac$ depends on $A$, the notation $flowcost_{(r,ac)}(A)$ is meaningful.)
Finally, a nodal cost metric has to be derived from $flowcost$; for example, the nodal cost can be defined as a weighted sum to reflect the aggregate performance of all source e2e-flows and the importance of $VO$ flows:
\begin{equation}
nodalcost_i(A)=\sum_{(r,ac)\in F\wedge s_r=i}\gamma_{ac}\times flowcost_{(r,ac)}(A),
\end{equation}
where $\gamma_{BE}=1$ and $\gamma_{VO} > 1$. It is convenient to normalize nodal costs to the all-neutral case:
\begin{equation}
\label{cost}
cost_i(A) = nodalcost_i(A) / nodalcost_i(\varnothing).
\end{equation}

For the above examples, under the flow-sparse and flow-dense traffic patterns, Table~\ref{tab:example-node-costs}a shows change of nodal costs between the all-neutral case (no TRAs) and after TRAs launched by the set of opportunistic attackers $A = \{1, 3, 8, 9\}$, as well as nodes' classification regarding the impact of the TRAs. Best-fit rank metric (\ref{rank}) with $\alpha=40$ and $\beta=10$, and $\gamma_{VO} = 2$ are assumed. 
Attackers (neutral nodes) whose costs have increased are classified as \emph{lose} (\emph{mind}) and the others as \emph{don't lose} (\emph{don't mind}).

\begin{table}[th]
  \small
  \centering
  \caption{Change of node costs and node classification after TRAs for $\alpha= 40$, $\beta= 10$, and 
$\gamma_{VO} = 2$.
}
\subfloat[$A = \{1,3,8,9\}$]{\label{tab:example-node-costs-a}
\begin{tabular}{@{}ccccc@{}}
\toprule
\multirow{2}{*}{Node} & \multicolumn{2}{c}{Flow-sparse}   & \multicolumn{2}{c}{Flow-dense}    \\
                      & Cost change  & State      & Cost change & State      \\
1                     & 133\%                  & \emph{lose}      & 11\%                   & \emph{lose}      \\
2                     & -40\%                  & \emph{don't mind} & -34\%                  & \emph{don't mind} \\
3                     & -40\%                  & \emph{don't lose} & -26\%                  & \emph{don't lose} \\
4                     & -49\%                  & \emph{don't mind} & -2\%                   & \emph{don't mind} \\
5                     & 104\%                  & \emph{mind}      & 2\%                    & \emph{mind}      \\
6                     & -49\%                  & \emph{don't mind} & 11\%                   & \emph{mind}      \\
7                     & 131\%                  & \emph{mind}      & -1\%                   & \emph{don't mind} \\
8                     & -41\%                  & \emph{don't lose} & -50\%                  & \emph{don't lose} \\
9                     & 89\%                   & \emph{lose}      & -28\%                  & \emph{don't lose} \\
10                    & -46\%                  & \emph{don't mind} & 8\%                    & \emph{mind}     \\ \bottomrule
\end{tabular}}\\
\subfloat[$A = \{1\dots10\}$]{\label{tab:example-node-costs-b}
\begin{tabular}{@{}ccccc@{}}
\toprule
\multirow{2}{*}{Node} & \multicolumn{2}{c}{Flow-sparse}   & \multicolumn{2}{c}{Flow-dense}    \\
                      & Cost change  & State      & Cost change & State      \\
1                     & 124\%           & \emph{lose}         & -8\%           & \emph{don't lose}    \\
2                     & -52\%           & \emph{don't lose}    & -34\%          & \emph{don't lose}    \\
3                     & 135\%           & \emph{lose}         & -37\%          & \emph{don't lose}    \\
4                     & -91\%           & \emph{don't lose}    & -5\%           & \emph{don't lose}    \\
5                     & 141\%           & \emph{lose}         & -34\%          & \emph{don't lose}    \\
6                     & -91\%           & \emph{don't lose}    & -36\%          & \emph{don't lose}    \\
7                     & 149\%           & \emph{lose}         & -6\%           & \emph{don't lose}    \\
8                     & -46\%           & \emph{don't lose}    & -7\%           & \emph{don't lose}    \\
9                     & 147\%           & \emph{lose}         & -20\%          & \emph{don't lose}    \\
10                    & -52\%           & \emph{don't lose}    & -24\%          & \emph{don't lose}  \\  \bottomrule
\end{tabular}}
\label{tab:example-node-costs}
\end{table}

One sees that, surprisingly, TRA behavior can both be harmful to an attacker and be harmless (even beneficial) to a node staying neutral. The reason is that from the viewpoint of an e2e-flow, TRAs may in various ways affect the number of competing \emph{VO} h-flows -- either decrease it (due to TRA$^-$) or increase it (reflecting stronger interference from nodes of the same route due to TRA$^+$).

\section{TRA Game}
\label{sec:game}
In a noncooperative game that arises, the nodes are players, $map_i$ is node $i$'s strategy, and \emph{cost} is the (negative) payoff function. 
A strategy profile $(map_i, i \in N)$ can be equivalently described as the set $A \subseteq N$ of opportunistic attackers.
A formal description of the game, of the form $\langle$player set, strategy profile set, payoff function$\rangle$, is therefore:
\begin{equation}
\langle N, 2^{N}, cost: 2^{N} \rightarrow \mathbf{R}^+\rangle.
\end{equation}

To investigate the nature of the game in a small-size MANET, $cost$ can be tabulated applying (\ref{OH})-(\ref{cost}) to all feasible strategy profiles. For a moderate-size MANET, various game scenarios can be analyzed by simulating the evolution (successive transitions) of the attacker set $A$.
Some interesting strategy profiles are: $\varnothing$ (all-neutral, corresponding to no TRAs being launched), and $N$ (corresponding to each node being an opportunistic attacker).
In the latter, any e2e-flow $(r, BE)$ experiences a TRA$^+$ at $s_r$, whereas any e2e-flow $(r, VO)$ with $|r| > 2$ experiences a TRA$^-$ at the first node in $r \setminus \{s_r, d_r\}$.

An observation that necessitates subtler game-theoretic treatment is that, contrary to the intuition behind $rank(\cdot)$ (whereby it is apparently beneficial to upgrade source traffic, as $\alpha > 0$, and to downgrade competing traffic, as $\beta > 1$), the TRA game is not a multiperson Prisoners' Dilemma (PD).
Specifically, TRA does not dominate neutral behavior, as seen from the presence of \emph{lose} nodes in the above examples.
Neither is TRA necessarily harmful to neutral nodes, as seen from the presence of \emph{don't mind} nodes.
This is because of the complex interplay of MAC contention, EDCA prioritization and intra-flow competition due to multi-hop forwarding in the presence of hidden stations (where packet transmissions from one station compete with those from up- and downstream stations one or two hops away).
Moreover, $A=\varnothing$ may, but need not be Pareto superior to $A=N$; in fact, for some traffic patterns (such as the flow-dense one in Table~\ref{tab:example-node-costs}b), the reverse is true, i.e., all the nodes are classified as \emph{don't lose}.


An important characterization of a game is through its \emph{Nash equilibria} (NE) \cite{Fudenberg1998}.
For $A \subseteq N$ and $i \in N$ define 
\begin{equation}
A^{[i]}=\left\{
\begin{aligned}
&A\setminus\{i\},&i\in A \\ 
&A \cup \{i\}, &i\notin A
\end{aligned}
\right. ;
\end{equation}
then $\hat A$ is a \emph{weak} (\emph{strict}) NE if and only if
\begin{equation}
\forall i \in N: cost_i(\hat A)\leq (<)  cost_i(\hat A^{[i]}).
\label{eq:weakNE}
\end{equation}
Numerical experiments show that the TRA game may possess multiple NE.
An exhaustive search of the set $2^{N}$ for 100 random MANET topologies with $|N| = 10$ and uniformly distributed route lengths $2\dots 5$ (with $\alpha$, $\beta$, and $\gamma_{VO}$ as in the above examples) reveals that among the $2^{|N|}$ feasible strategy profiles, typically up to $5\%$ are NE for the flow-sparse, and below $1\%$ are NE for the flow-dense traffic pattern, cf. the x-coordinates of the dots in Fig.~\ref{fig:TRAgame-NEhits-sparse} and~\ref{fig:TRAgame-NEhits-dense}.
The vast majority of NE are weak.

Given the non-PD nature of the TRA game, one needs to establish conditions under which TRAs pose a real danger and so defense is necessary. 
Hence, one needs to predict nodal costs at the strategy profile that nodes will arrive at under some model of rational play. Note that while rational play leads to an NE \cite{Kalai1993}, MANET nodes are better modeled as \emph{boundedly rational}, i.e., exhibiting limited complexity, perseverance, and foresight. Such nodes can be reasonably expected to adopt a simple multistage attack strategy that leads to some form of equilibrium, cf. \cite{Konorski2014,Ho1996}.

\section{Multistage TRA Game}
\label{sec:multistage-game}
Suppose the TRA game is played in stages $k = 1, 2, {\dots}$, and in each stage a node's behavior is either to \emph{attack} all e2e-flows it can attack (i.e., become an opportunistic attacker) or to stay \emph{neutral}.
Let $A(k) \subseteq N$ be the set of attackers in stage $k$.
Of interest is the evolution (in particular asymptotic behavior) of $A(k)$, starting from any $A(0) \subseteq N$, under some multistage attack strategy that can be justified as rational in some sense.
A possible heuristic multistage attack strategy with a rational trait is for a node to:
\begin{enumerate}
\item disallow a behavior change if the current cost is the smallest over a predefined number of recent stages (referred to as \emph{cost memory}, $CM$); the node is then called \emph{satisfied}, 
\item if a behavior change is allowed (node is currently dissatisfied), decide it with a probability that depends on the history of own play; the change from the current behavior is driven by the excess of past stages where the same behavior led to a cost increase over those where it did not.
\end{enumerate}
The above multistage strategy can be formalized simply by specifying how $A(k)$ arises from $A(k-1)$. This is given by Algorithm \ref{algorithm}, where $\varepsilon_\pi$ denotes a random event occurring with probability $\pi$, $\sigma : R \rightarrow [0, 1]$ is a nondecreasing function with $\lim_{x\rightarrow-\infty}\sigma(x) = 0$ and $\lim_{x\rightarrow+\infty}\sigma(x) = 1$, and the logical condition $\varphi_i(k)$ expresses node $i$'s satisfaction in stage $k$:
\begin{equation}
\label{satisf}
\varphi_i(k)=\left\{ 
\begin{aligned}
&\text{false}, k<CM \\ 
&\forall_{l=1\dots CM}cost_i(A(k))\le cost_i(A(k-l)), k\ge CM
\end{aligned}
\right.
\end{equation}
\begin{algorithm}[t]
\small
\SetAlgoLined
\KwData{Stage $k$}
\KwResult{Set of attackers $A(k)$}
  \eIf{$k=0$}{
   initialize $Lose$, $DontLose$, $Mind$, and $DontMind$ counters\;
   $A(k) \gets$ any subset of $N$\;
   }{
   $A(k) \gets \{i\in A(k-1):\varphi _i(k-1)\vee \neg \varepsilon_{\sigma({Lose}_i-{DontLose}_i)}\}\cup \{i\notin A(k-1):\neg \varphi_i(k-1)\wedge \varepsilon_{\sigma ({Mind}_i-{DontMind}_i)}\}$\;
	\For{$i \in N$}{
		\eIf{$cost_i(A(k)) > cost_i(A(k-1))$}{
			\uIf{$i \in A(k-1)$}{
				increment $Lose_i$
			}
			\uElse{
				increment $Mind_i$
			}
		}
		{
        	\eIf{$i \in A(k-1)$}{
			increment $DontLose_i$
		}
		{
			increment $DontMind_i$		
		}}
	}
  }
\caption{Multistage attack strategy}
\label{algorithm}
\end{algorithm}
Note that the first $CM$ stages make an exploratory warm-up, where all nodes are considered dissatisfied and so TRAs are launched at random, governed by $\varepsilon_\pi$.

\section{Multistage Game Analysis}

Starting from some fixed $A(0)$, the evolution of $A(k)$ and corresponding normalized node costs under the above multistage attack strategy can be easily recreated in Monte Carlo simulations.
The simulations were carried out for the above MANET examples, with $CM = 10$, $\sigma(x) = 1/(1 + e^{-x})$ (the sigmoid function) and the other parameters ($\alpha$, $\beta$, $\gamma_{VO}$) are set as in the previous examples.

\begin{figure}
\centering
\includegraphics[width=0.95\columnwidth]{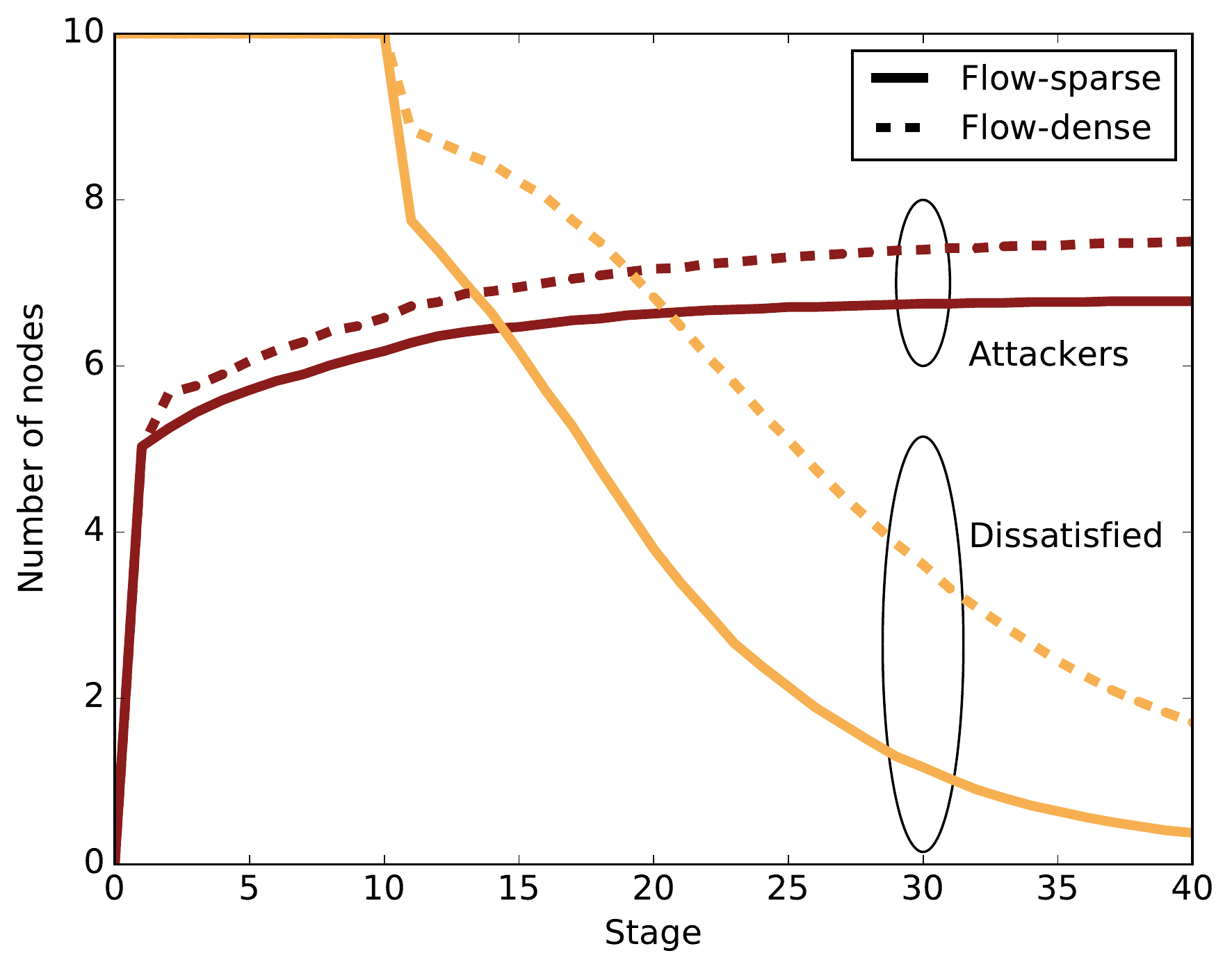}
\caption{Evolution of multistage TRA game for flow-sparse and flow-dense traffic patterns: number of attackers and of dissatisfied nodes.}
\label{fig:TRAgame-multistage}
\end{figure}

For the flow-sparse and flow-dense traffic patterns, Fig.~\ref{fig:TRAgame-multistage} plots against $k$ the following characteristics, averaged over 100 random MANET instances with $|N| = 10$ and uniformly distributed route lengths $2\dots5$, and 100 runs per instance with fixed $A(0)=\varnothing$ ($\alpha$, $\beta$, and $\gamma_{VO}$ are set as before):
\begin{itemize}
\item number of attackers, i.e., $|A(k)|$,
\item number of dissatisfied nodes, i.e., $\{i\in N:\neg \varphi _i(k)\}|$.
\end{itemize}

In randomly generated MANET instances, $A(k)$ typically (in 97.3\% of flow-sparse runs and 93.3\% of flow-dense runs) converges over time to a ``quasi-equilibrium'' $A_\infty$ such that $\varnothing \neq A_\infty \neq N$.\footnote{The notion of ``quasi-equilibrium" we employ reflects some existing approaches to bounded rationality listed in \cite{Ho1996}: the limited cost memory $CM$ implies a node's constrained complexity and resources, and the form of (\ref{satisf}) implies limited perseverance and a myopic attitude.}
$A_\infty$ may differ from run to run on account of the randomness inherent in Algorithm \ref{algorithm}, and may also  depend on $A(0)$. 
However, averaged over multiple runs, $|A_\infty|$ seems insensitive to $A(0)$: similar plots were produced when $A(0)$ was randomly chosen in successive runs.
Moreover, $cost_i(A_\infty) < 1$ is observed for some or all $i \notin A_\infty$.
This confirms that at a ``quasi-equilibrium'', TRAs can be harmless to some neutral nodes.

Based on the plots one also conjectures that the convergence occurs regardless of scenario (albeit may be slow for large $CM$), i.e., all the nodes eventually become satisfied.
The intuitive explanation is that nodes whose costs have not increased recently do not leave the satisfied set, and the others are more likely to try different behavior and so to lower their costs in the near future (indeed, observed nodes' costs cease to increase from some stage on).

An interesting characterization of the observed $A_\infty$, reflecting the rationality of Algorithm \ref{algorithm}, is that there are eventually very few nodes $i$ for which $cost_i(A_\infty^{[i]}) < cost_i(A_\infty)$, and which also fulfill the following:
\begin{itemize}
\item  if $i\in A_\infty$ then $\forall_{j \in A_\infty^{[i]}}cost_j(A_\infty^{[i]})\geq cost_j(A_\infty)$, i.e., node $i$ is an attacker, but if had stayed neutral, would have decreased its cost without decreasing that of any other attacker (hence, without bolstering other attackers' satisfaction), 
\item  if $i\notin A_\infty$ then $\forall _{j\notin A_\infty^{[i]}} cost_j(A_\infty^{[i]})\leq cost_j(A_\infty)$, i.e., node $i$ is neutral, but if had attacked, would have decreased its cost without increasing that of any other neutral node (hence, without provoking neutral nodes' dissatisfaction, which might lead to more TRAs).
\end{itemize}
In the simulations, the former type of node $i$ was almost never observed, and the latter averaged around $6\%$ and $4.5\%$ of $|N|$ in the flow-sparse and flow-dense runs, respectively.

\begin{figure*}
  \vspace{0.01in}
\centering
\subfloat[Percentage of instances an NE is asymptotically arrived at, flow-sparse traffic pattern.]{\includegraphics[width=0.32\textwidth]{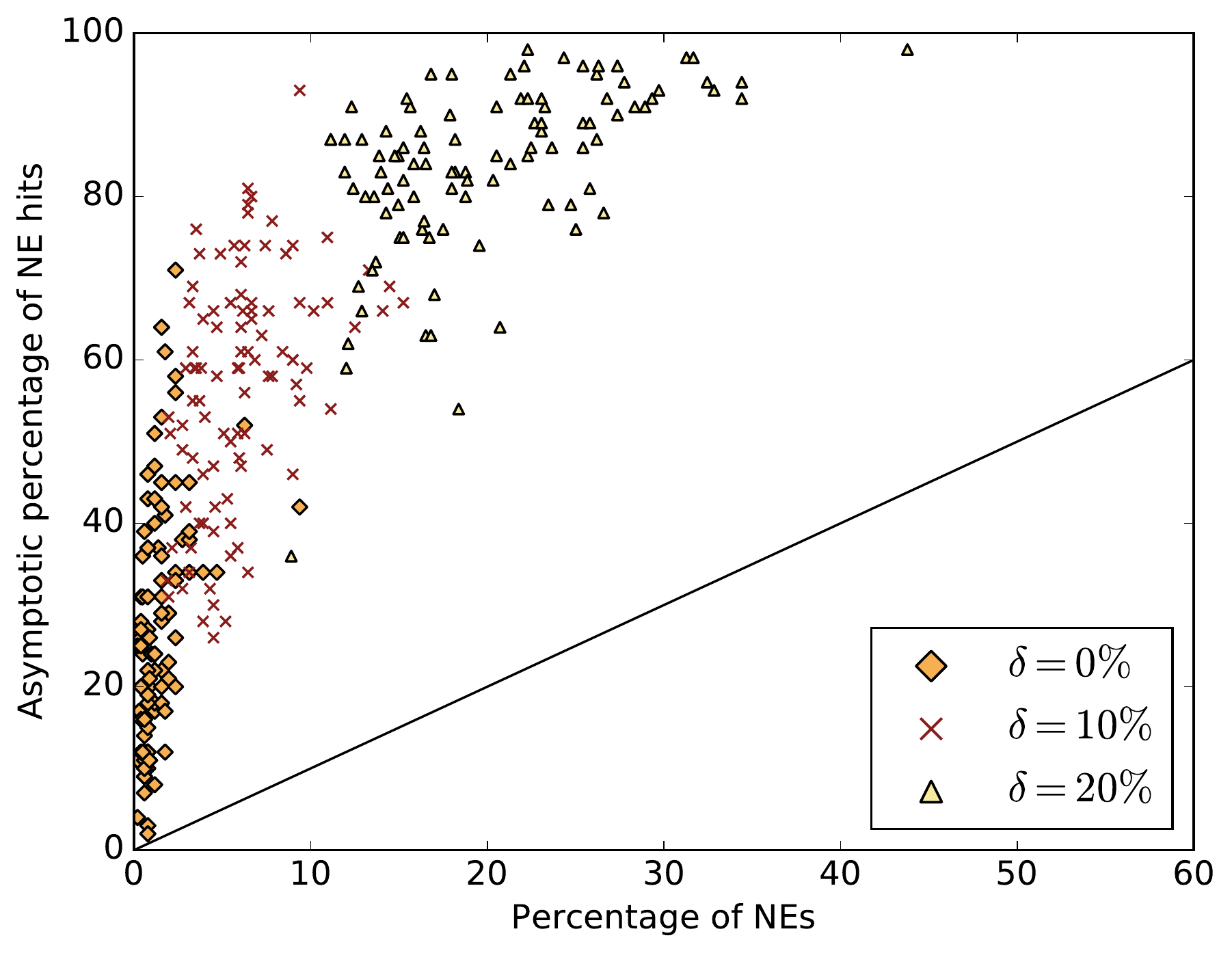}%
\label{fig:TRAgame-NEhits-sparse}}
\hfill
\subfloat[Percentage of instances an NE is asymptotically arrived at, flow-dense traffic pattern.]{\includegraphics[width=0.32\textwidth]{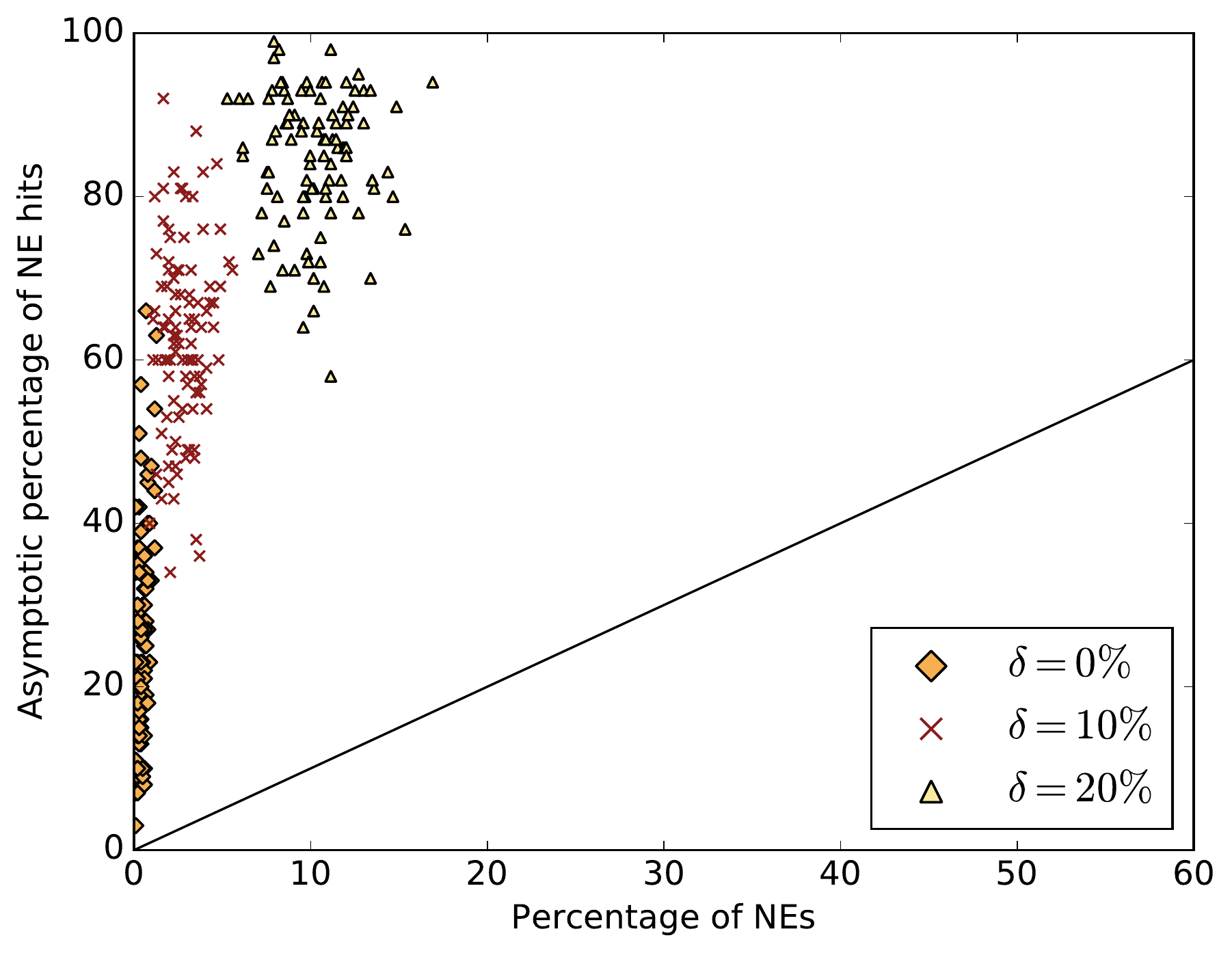}%
\label{fig:TRAgame-NEhits-dense}}
\hfill
\subfloat[Effectiveness]{\includegraphics[width=0.32\textwidth]{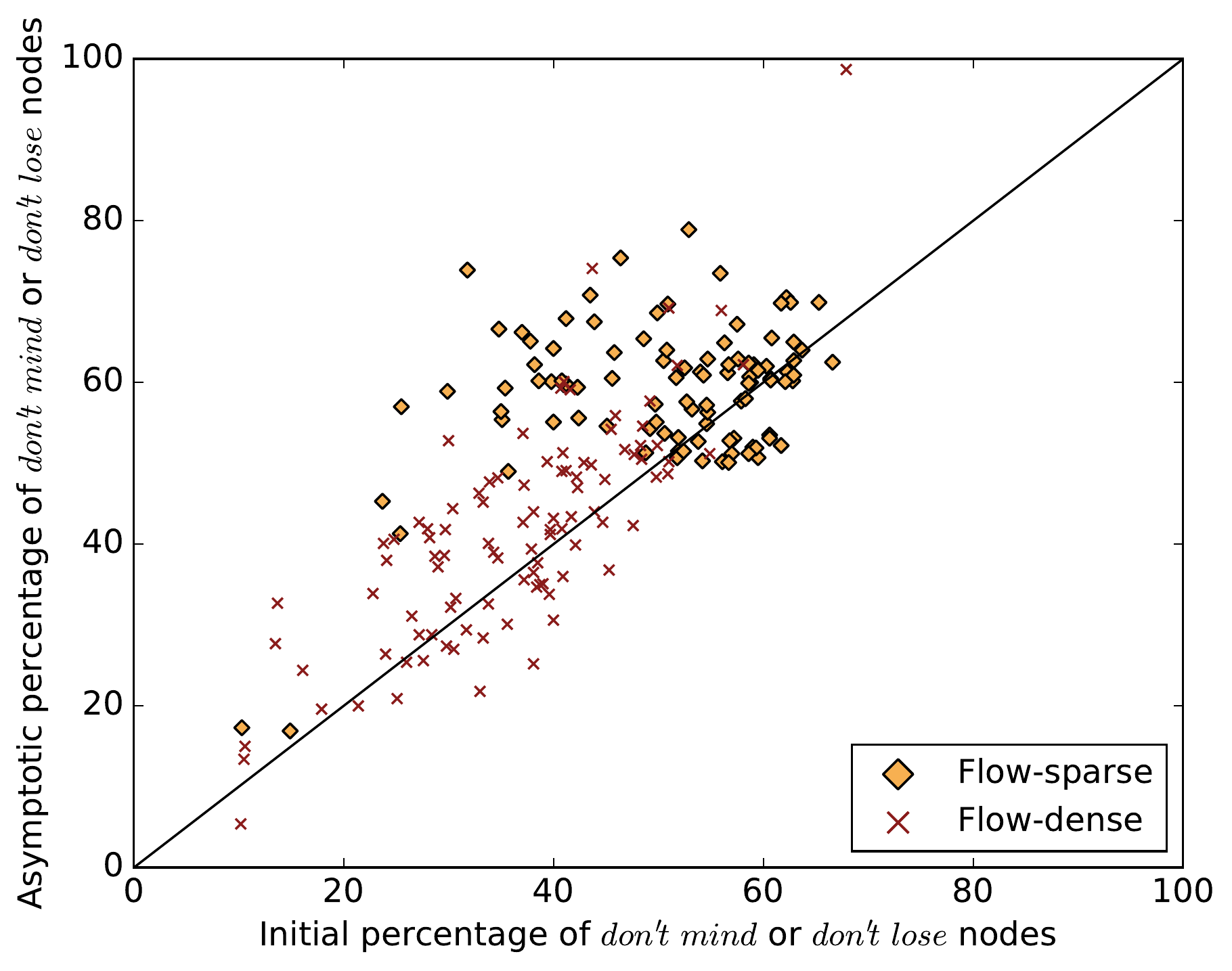}%
\label{fig:TRAgame-effectiveness}}
\caption{Performance of the multistage attack strategy given by Algorithm \ref{algorithm} and its statistical significance. If a MANET instance produces a dot above the $y=x$ line then the asymptotic ``NE hits'' and improvement of the percentage of beneficiaries are larger than accidental.}
\label{fig:TRAgame-performance}
\vspace{-0.5cm}
\end{figure*}

One can also judge the rationality of Algorithm \ref{algorithm} by ``NE hits'' (the percentage of runs where an NE is asymptotically arrived at) in relation to the proportion of NE among the $2^{|N|}$ feasible strategy profiles.
If the former is distinctly larger than the latter, the NE reached by the multistage strategy is not ``accidental'', and the strategy can be said to be \emph{NE-seeking}.
To get a more comprehensive view, one can extend the notion of NE to $\delta$-NE, where up to a fraction $\delta$ of the inequalities (\ref{eq:weakNE}) are violated.
The plots in Fig.~\ref{fig:TRAgame-NEhits-sparse} and \ref{fig:TRAgame-NEhits-dense} have been obtained for the same 100 random MANET instances as before, with $\delta=0$, 10\%, and 20\%.
Each dot corresponds to a MANET instance with ``NE hits'' obtained from 100 runs with random $A(0)$.
For both flow-sparse and flow-dense traffic patterns the NE-seeking property is visible.

A good characterization of the above multistage strategy is its \emph{effectiveness} reflecting the ability to improve the perceived QoS over successive stages. Assuming that all legitimate QoS requirements are satisfied at the all-neutral strategy profile $\varnothing$, of interest is the average percentage of asymptotic beneficiary (\emph{don't mind} or \emph{don't lose}) nodes against the analogous average percentage for corresponding initial strategy profiles.
If the former percentage is larger, the multistage strategy can be said to be asymptotically effective.
The plot in Fig.~\ref{fig:TRAgame-effectiveness} reflects the same 100 MANET instances, each producing a dot obtained by averaging over 100 runs with random $A(0)$.
Our multistage attack strategy turns out asymptotically effective in more than 70\% of MANET instances both for the flow-sparse and flow-dense traffic patterns. Hence, the asymptotic outcomes typically feature more beneficiary nodes than do the corresponding initial strategy profiles.
One concludes that Algorithm 1 expresses selfish nodes' bounded rationality and yields non-selfish nodes an effective defense against TRAs by responding in kind. The fact that not all the nodes end up as attackers, but all are satisfied with the costs, suggests that harmful TRAs are curbed in the first place, whereas harmless ones need not be. It also shows that, fortunately, responding in kind is not a punishment that leads to a spiral of ``punishing the punishers'' \cite{Banchs2013}.




\section{Conclusion}
\label{sec:conclusions}
We have proposed a formal model of TRAs for multi-hop ad hoc networks.
This allowed us to rigorously define the arising TRA game, for which a multistage attack strategy has been proposed. The strategy is arguably boundedly rational, at the same time can serve as an effective respond-in-kind countermeasure to TRAs.

The analysis of alternative boundedly rational strategies (e.g., reinforcement learning, trial-and-error, regret-based), as well as rigorous proofs of convergence are left for future research.
Also, more work should be devoted to \emph{systematically} prevent or curb only harmful TRAs, as was done in a WLAN setting \cite{Konorski2014}. Finally, simulations of Algorithm \ref{algorithm} in realistic wireless networks should verify our performance model as well as reveal the impact of transmission impairments, traffic volume, and end-to-end protocols such as TCP.

\section*{Acknowledgments}

The work of Jerzy Konorski is supported by the Statutory Fund 
of the Faculty of
Electronics, Telecommunications and Informatics, Gdansk University of 
Technology. 
The work of Szymon Szott is supported by the AGH University of Science and Technology (contract no. 11.11.230.018).
Szymon Szott also received a conference travel grant from the National Science Centre, Poland (reg. no. 2017/01/X/ST7/00158).



\bibliographystyle{IEEEtran}
\bibliography{tra-game}

\end{document}